\documentclass[pra,twocolumn,amsmath,amssymb]{revtex4-1}
\usepackage{hyperref}
\pdfoutput=1
\usepackage{graphicx}
\usepackage{epsfig}
\usepackage{epstopdf}

\usepackage{amsmath}
\usepackage{bm}
\usepackage{amssymb}
\usepackage{textcomp}
\usepackage{stmaryrd} 
\usepackage{ulem}

\usepackage{import}
\usepackage{color}
\usepackage{verbatim}

\usepackage{tabularx}
\usepackage{nccmath}

\usepackage{float}

\newcommand\abs[1]{\left| #1 \right|}

\def\XXint#1#2#3{{\setbox0=\hbox{$#1{#2#3}{\int}$}
     \vcenter{\hbox{$#2#3$}}\kern-.5\wd0}}

\begin{document}

\title{Monitoring Electrostatically-Induced Deflection, Strain and Doping in Suspended Graphene using Raman Spectroscopy}

\author{Dominik Metten}
\affiliation{Institut de Physique et Chimie des Mat\'eriaux de Strasbourg and NIE, UMR 7504, Universit\'e de Strasbourg and CNRS, 23 rue du L\oe{}ss, BP43, 67034 Strasbourg Cedex 2, France}

\author{Guillaume Froehlicher}
\affiliation{Institut de Physique et Chimie des Mat\'eriaux de Strasbourg and NIE, UMR 7504, Universit\'e de Strasbourg and CNRS, 23 rue du L\oe{}ss, BP43, 67034 Strasbourg Cedex 2, France}

\author{St\'ephane Berciaud}
\email{stephane.berciaud@ipcms.unistra.fr}
\affiliation{Institut de Physique et Chimie des Mat\'eriaux de Strasbourg and NIE, UMR 7504, Universit\'e de Strasbourg and CNRS, 23 rue du L\oe{}ss, BP43, 67034 Strasbourg Cedex 2, France}


\begin{abstract}
Electrostatic gating offers elegant ways to simultaneously strain and dope atomically thin membranes. Here, we report on a detailed \textit{in situ} Raman scattering study on graphene, suspended over a Si/SiO$_2$ substrate. In such a layered structure, the intensity of the Raman G- and 2D-mode features of graphene are strongly modulated by optical interference effects and allow an accurate determination of the electrostatically-induced membrane deflection, up to irreversible collapse. The membrane deflection is successfully described by an electromechanical model, which we also use to provide useful guidelines for device engineering. In addition, electrostatically-induced tensile strain is determined by examining the softening of the Raman features. Due to a small residual charge inhomogeneity $\sim 5\times 10^{10}~\rm cm^{-2}$, we find that non-adiabatic anomalous phonon softening is negligible compared to strain-induced phonon softening. These results open perspectives for innovative Raman scattering-based readout schemes in two-dimensional nanoresonators.



\textbf{Keywords:} {suspended graphene, two-dimensional materials, Raman spectroscopy, strain, doping, optical interference, NEMS.}
\end{abstract}


\maketitle

\noindent\textbf{Introduction} 
Electrostatic gating is one of the most commonly employed actuation schemes in nanomechanical resonators~\cite{Ekinci2005}. In particular, field-effect transistor geometries have been adapted to fabricate nano-electromechanical resonators using individual carbon nanotubes~\cite{Sazonova2004}, graphene~\cite{Bunch2007,Chen2009} and, more recently, atomically thin transition metal dichalcogenides~\cite{Lee2013,Castellanos-Gomez2013}. In such devices, the ultimate thinness of the suspended nanoresonator leads to high electromechanical susceptibility and possible coupling between electrostatically-induced strain and doping. As a classic example, similar designs of suspended graphene transistors have been used not only to fabricate electro-mechanical~\cite{Chen2009} or opto-electromechanical devices~\cite{Bunch2007,Barton2012,Reserbat2012} but also to probe quasi-ballistic electron transport~\cite{Bolotin2008,Du2008} and many-body effects~\cite{Elias2011,Faugeras2015} in the vicinity of graphene's charge neutrality (Dirac) point. Electrostatic pressure can thus intentionally be applied to control the position, shape, and motion of an atomically thin suspended membrane but may also be an undesired side-effect when examining intrinsic transport properties~\cite{Huang2011,Zhang2014,Medvedyeva2011}, ultimately leading to irreversible collapse~\cite{Bolotin2008,Bao2012}. As a result, \textit{in situ} probes are needed in order to examine the subtle interplay between doping, electron transport, motion and strain in  electrostatically-actuated membranes. 

Here, we employ micro-Raman scattering spectroscopy as a minimally invasive and highly accurate technique  to simultaneously monitor electrostatically-induced deflection, strain and doping in a pristine suspended graphene monolayer. Our analysis is based on the large changes in intensity, frequency, and linewidth of the main Raman features of graphene subjected to an electrostatic pressure. The measured deflection is well-captured by an electromechanical model~\cite{Landau1970,Medvedyeva2011,Bao2012}, which solely uses the built-in tension in graphene as a fitting parameter. Our model accurately predicts the critical gate bias and highest  doping level achievable before device collapse, therefore providing useful  guidelines for opto-electromechanical device engineering. Here, the charge carrier density remains significantly below $10^{12}\:\rm cm^{-2}$, with a small residual charge inhomogeneity of $\sim 5\times 10^{10}~\rm cm^{-2}$. In these conditions, the Raman features of electrostatically-gated suspended graphene are chiefly affected by strain-induced phonon softening, with only minor contributions from non-adiabatic electron-phonon coupling~\cite{Lazzeri2006,Ando2006,Yan2008}.

\begin{figure*}[!t]
\begin{center}
\includegraphics[width=1\linewidth]{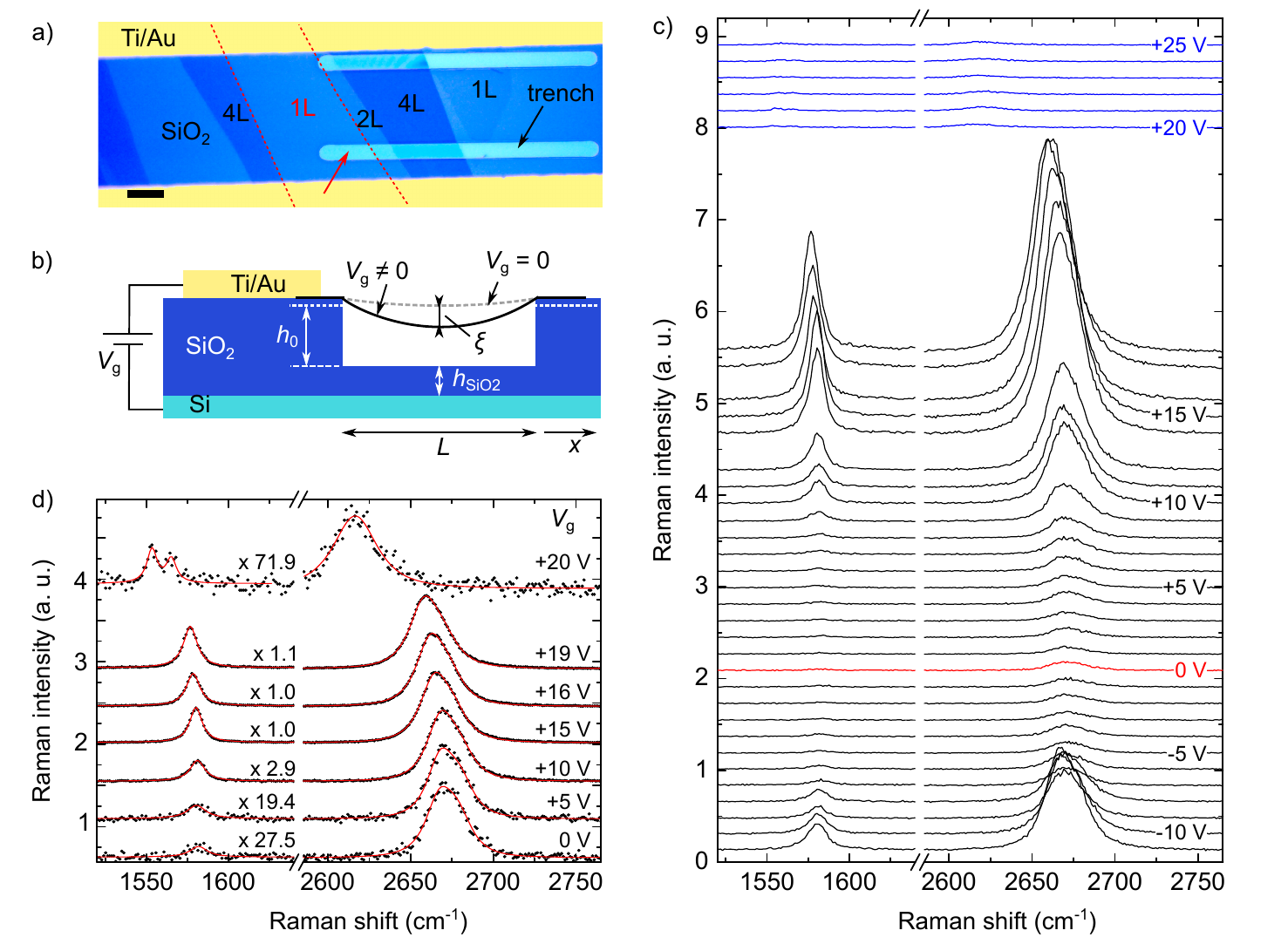}
\caption{(a) Optical micrograph of a suspended graphene device. The red dashed line marks the limit of the graphene monolayer (1L), which is neighbored by a bi- and tetralayer. The scale bar is $10$\,$\mu$m. (b) Schematic side view of the device. (c) Cascade plot of Raman spectra recorded in the middle of the suspended graphene membrane, as a function of $V_\mathrm{g}$. Spectra are recorded at $4$\,K. The spectrum at $V_\mathrm{g}=0$ is drawn in red. Spectra recorded after sample collapse ($V_\mathrm{g}>+19$\,V) are shown in blue. (d) Selected Raman spectra for different $V_\mathrm{g}$, normalized with respect to the integrated intensity of the 2D-mode feature. The red solid lines are fits to the G- (2D-) mode features using Lorentzian (modified Lorentzian~\cite{Basko2008,Berciaud2013}) profiles.}
\label{fig1}
\end{center}
\end{figure*}

\noindent\textbf{Methods} 
Our typical sample consists of a mechanically exfoliated graphene monolayer suspended over a $4.9$\,$\mu$m-wide trench pre-patterned by optical lithography and reactive ion etching on a Si/SiO$_2$ substrate (oxide thickness of $500$\,nm, of which $h_{\rm SiO_2}=212\pm5$\,nm of SiO$_2$ are left)~\cite{Berciaud2009}. The suspended monolayers (1LG) are identified via optical microscopy and Raman spectroscopy as discussed below. To avoid contamination with resist and solvents, Ti/Au contacts are evaporated through a transmission electron microscopy  grid, used as a shadow mask (see Fig.~\ref{fig1}a,b). Finally, the gold contact pads are wire-bonded and the sample is cooled down to $4$\,K using an He flow optical cryostat. Micro-Raman spectra are recorded as a function of the back-gate voltage $V_\mathrm{g}$ in back-scattering geometry using a laser beam with a wavelength of $532$\,nm focused onto a $\approx\:1.2\rm \mu m$-diameter spot onto the sample using a $50\times$ objective with a numerical aperture of $0.65$. The laser power impinging on the sample is maintained below $300$\,$\mu$W to avoid photothermally-induced changes in the Raman features and sample damage. In the following, we focus on the Raman G and 2D modes, which involve zone-center and near zone-edge optical phonons, respectively~\cite{Ferrari2013}. The defect-allowed D-mode has a negligible intensity and is not considered here. The peak frequency $\omega_\mathrm{G}$ ($\omega_\mathrm{2D}$), full-width at half maximum $\Gamma_\mathrm{G}$ ($\Gamma_\mathrm{2D}$), and integrated intensity $I_\mathrm{G}$ ($I_\mathrm{2D}$) of the G-mode and 2D-mode features are extracted from single Lorentzian and modified Lorentzian fits~\cite{Basko2008,Berciaud2013}, respectively. As previously demonstrated~\cite{Bolotin2008,Du2008,Elias2011,Berciaud2009,Ni2009,Berciaud2013,Metten2013} and also shown in the Supporting Information, suspended graphene exhibits low residual charge carrier density and inhomogeneity, both below $10^{11}$\,cm$^{-2}$. Consequently, we assume that graphene is quasi neutral at $V_{\rm g}=0~\rm V$. We note $h_0$, the initial distance between the membrane and the underlying SiO$_2$ layer at $V_{\rm g}=0$ and $\xi$ the gate-induced deflection in the middle of the trench (see Fig.~\ref{fig1}b).

\begin{figure*}[!t]
\begin{center}
\includegraphics[width=\linewidth]{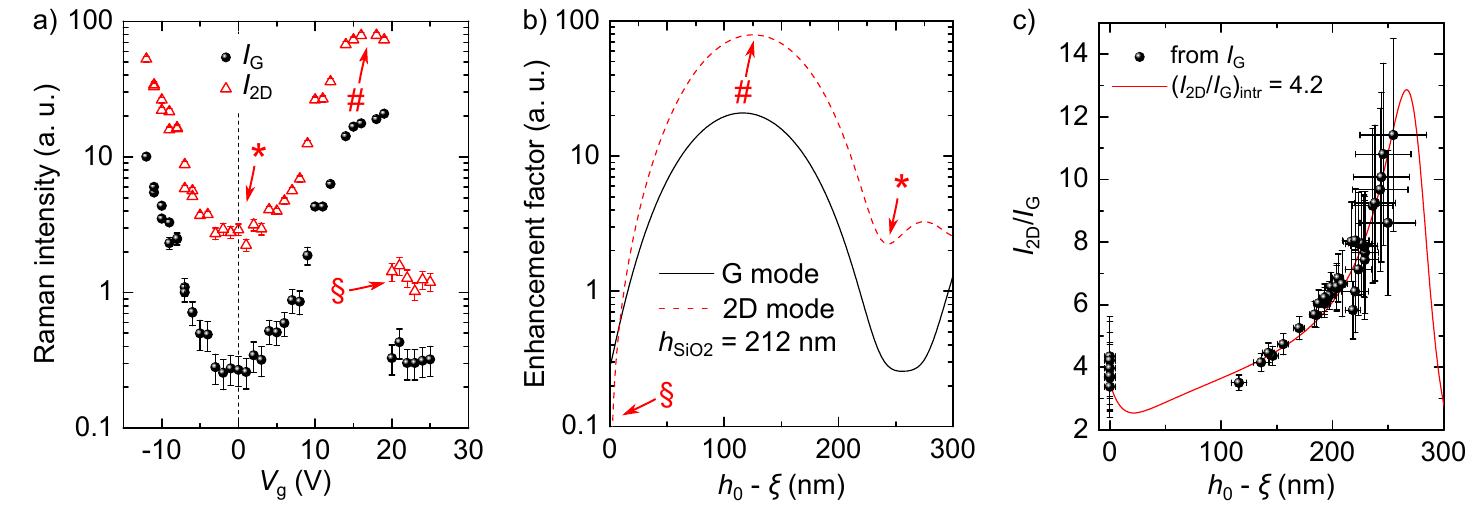}
\caption{(a) Semilogarithmic plot of $I_\mathrm{G}$ (filled black squares) and $I_\mathrm{2D}$ (open red triangles) as a function of $V_\mathrm{g}$. (b) Raman enhancement factors as a function of $h_0-\xi$, the distance between graphene and the SiO$_2$ layer, calculated for $h_\mathrm{SiO_2}=212$\,nm and $\lambda_\mathrm{laser}=532$\,nm. The calculated Raman intensities at heights marked with star, hash and paragraph symbols in (b) can be identified in (a). (c) Evolution of $I_\mathrm{2D}/I_\mathrm{G}$ as a function of $h_0-\xi$, determined from the evolution of $I_\mathrm{G}$ with $V_\mathrm{g}$.  The red line is a fit based on a multiple interference model using  an \textit{interference-free} reference value of $(I_\mathrm{2D}/I_\mathrm{G})_\mathrm{intr}=4.2$.}
\label{fig2}
\end{center}
\end{figure*}
 
\noindent\textbf{Electrostatically-induced deflection} 
Figure~\ref{fig1}c shows a cascade plot of Raman spectra recorded during a gate bias sweep from $V_\mathrm{g}=-11~\rm V$ to $V_\mathrm{g}=+25~\rm V$. Several interesting features can be observed as $V_\mathrm{g}$ rises up to $+19$\,V.  First, the integrated intensity of the Raman features increases strikingly. Second, as can be more clearly seen in Fig.~\ref{fig1}d, the integrated intensity ratio of the 2D- and G-mode features, $I_\mathrm{2D}/I_\mathrm{G}$, augments significantly. Third, the Raman G- and 2D-mode features downshift. The first and second observations stem from optical interference effects, which have a drastic impact on the Raman intensities in graphene devices within a multilayered structure~\cite{Yoon2009,Reserbat2013,Metten2014}. Indeed, the applied gate bias induces an electrostatic pressure that pulls the graphene layer towards the underlying SiO$_2$ layer, leading to gate-dependent Raman enhancement factors. For a quantitative analysis, $I_\mathrm{G}$ and $I_\mathrm{2D}$ are plotted as a function of $V_\mathrm{g}$ in Fig.~\ref{fig2}a (this plot includes a gate sweep from $V_{\rm g}=0$ to $-12$\,V and from $-12$ to $+25$\,V), along with, in Fig.~\ref{fig2}b, the Raman enhancement factors for the G- and 2D-mode features, computed using a multiple interference model~\cite{Yoon2009,Metten2014}. From this model, we are able to estimate the central deflection $\xi$. 

Here, $I_\mathrm{G}$ and $I_\mathrm{2D}$ are minimal for $V_\mathrm{g}=0$ and, as expected, show nearly identical behaviors at positive and negative $V_\mathrm{g}$. Interestingly, $I_\mathrm{2D}$ increases from its minimal value at $V_\mathrm{g}\approx 0$ to a maximum at $V_\mathrm{g}=+17$\:V, which we assign to an initial height $h_0\approx 250 \pm 20 \rm~ nm$ and to a deflection $\xi=134\pm  5~ \rm nm$, respectively. We note that $h_0$ is a little smaller than the actual depth of the trench ($288\pm5~\rm nm$), implying that the graphene membrane is slightly concave. This situation may be attributed to the mismatch between the thermal expansion  coefficients of the Si/SiO$_2$ substrate and of graphene~\cite{Bao2009,Yoon2011a}, as well as to the native slack of the as-exfoliated sample and adhesion to the sidewalls of the trench~\cite{Chen2009,Singh2010,Bunch2008,Kitt2013,Metten2014}. Since the Raman-scattered G- and 2D-mode photons have different wavelengths, the measured ratio $I_\mathrm{2D}/I_\mathrm{G}$, also depends on $\xi$ (see Fig.~\ref{fig2}c)~\cite{Yoon2009,Metten2015} and is in very good agreement with the calculated Raman enhancement factors~\footnote{In the range of gate biases applied here, $I_\mathrm{G}$ and $I_\mathrm{2D}$ are expected to vary marginally due to electrostatically-induced doping \cite{Yan2007,Froehlicher2015}} using an interference-free intrinsic ratio $\left(I_\mathrm{2D}/I_\mathrm{G}\right)_{\rm intr}$ of $4.2\pm 0.5$, consistent with our previous findings~\cite{Metten2015}.

The sudden drop in the Raman intensities above $V_\mathrm{g}=+19$\,V (see Fig.~\ref{fig1}c-d and~\ref{fig2}a) is inconsistent with a continuous increase of $\xi$ and is a strong indication that the membrane has collapsed and now adheres strongly to the underlying SiO$_2$ layer. Indeed, the Raman intensity is expected to be very low for the layered system Si/SiO$_2$(212\,nm)/graphene/vacuum  when $\xi\rightarrow h_0$ (see Fig.~\ref{fig2}b). The drop in the Raman intensities is accompanied by a downshift of the G- and 2D-mode features and by a splitting of the G-mode feature, characteristic of sizable uniaxial strain~\cite{Mohiuddin2009,Huang2009} (see Fig.~\ref{fig1}d). These observations suggest that the membrane remains attached on the edges of the trench, as confirmed by scanning electron microscopy (SEM) imaging (see Supporting Information). The G- and 2D-mode downshifts allow us to estimate a uniaxial strain of $\sim 1\%$~\cite{Mohiuddin2009}, that is consistent with a rough estimation based on the SEM image.

\begin{figure*}[!t]
\begin{center}
\includegraphics[width=\linewidth]{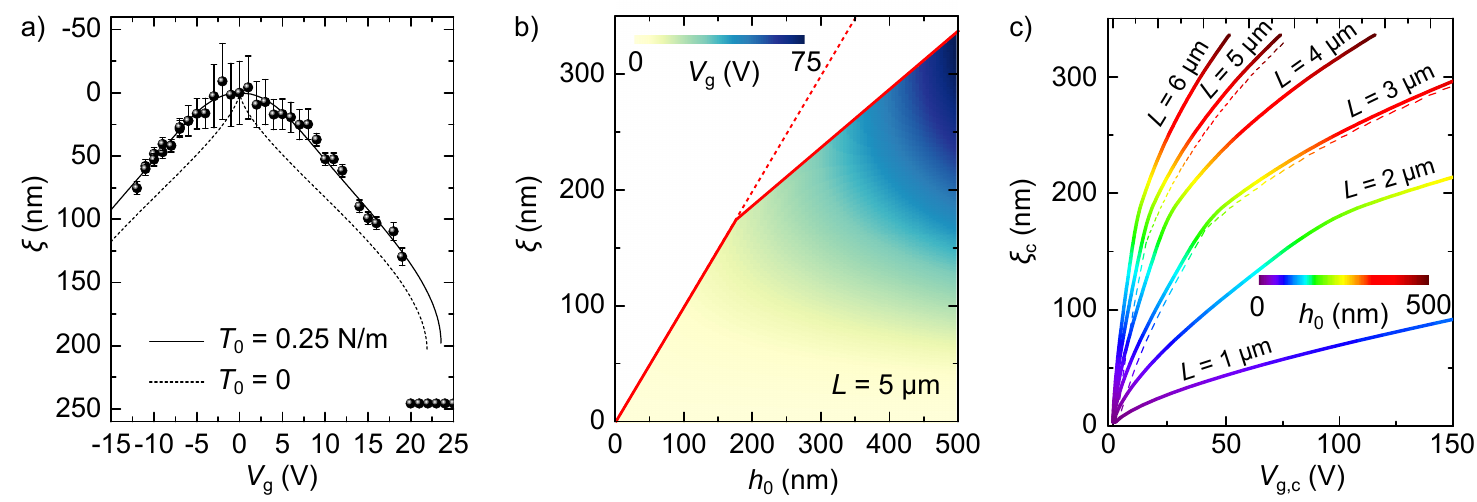}
\caption{(a) Graphene deflection $\xi$ in the middle of the trench, determined from the variation of $I_\mathrm{G}$, as a function of $V_\mathrm{g}$. The dashed (solid) line is a fit obtained by solving Eq.~\eqref{eq:pz} with $T_0=0$ ($T_0=0.25$\,N/m). (b) False color map of the gate voltage required to attain a given $\xi$ as a function of $h_0$, for $L=5$\,$\mu$m and an initial SiO$_2$ thickness of 500~nm (see Fig.~\ref{fig1}b).  (c) Critical deflection $\xi_{\rm c}$ as a function of the corresponding critical gate voltage $V_{\rm g,c}$. The data are extracted from color plots, as in (b), by going along the border highlighted with a solid red line in (b). The critical (straight) lines are shown for different trench widths $L$. The color code indicates the corresponding $h_{0}$. For $L=3$ and $5$\,$\mu$m, the critical (dashed) lines are added for a finite $T_0=0.3~\rm N/m$.}
\label{fig3}
\end{center}
\end{figure*}

Fig.~\ref{fig3}a displays $\xi$ as a function of $V_\mathrm{g}$, extracted from the variations of $I_{\rm G}$. Very similar results are obtained from the analysis of $I_{\rm 2D}$ (see Supporting Information). The large error bars near $V_\mathrm{g}=0$ account for the broad valley in the interference pattern (see Fig.~\ref{fig2}b). From the theory of elasticity~\cite{Medvedyeva2011,Bao2012}, $\xi$ is connected to a (here electrostatic) pressure load $p_{\rm el}$ through
\begin{equation}
p_\mathrm{el}(\xi)=\frac{64Et}{3(1-\nu^2)L^4}\,\xi^3+\frac{8T_0}{L^2}\,\xi,
	\label{eq:pz}
\end{equation}
where $E$ is the Young's modulus, $\nu$ the Poisson ratio, $t$ the thickness of the membrane, $T_0$ its built-in tension, and $L$ the trench width (see Fig.~\ref{fig1}b). The scaling $p_\mathrm{el}\propto\xi^3$ is well-known for membranes (with no bending rigidity)~\cite{Hencky1915,Komaragiri2005,Yue2012}, and $T_0$ can be interpreted as an effective bending rigidity moderating the deflection, as $p_\mathrm{el}$ scales linearly with $\xi$ for thin plates~\cite{Landau1970,Vlassak1992,Beams1995,Williams1997,Medvedyeva2011}. Eq.~\eqref{eq:pz} is obtained by assuming translational invariance along the trench and a parabolic membrane profile perpendicular to the trench ($x$ direction), which is a first approximation, because back-coupling mechanisms between deflection and charge redistribution may affect the membrane profile~\cite{Medvedyeva2011}. Noteworthy, Eq.~\eqref{eq:pz} also involves a uniform pressure load over the membrane, which is not the case in our experiment since $\xi$ attains values that are on the same order of magnitude as $h_{0}$. However, we can model our experimental data by considering an effective value of $p_\mathrm{el}$, obtained by averaging the local capacitance $c(x)$ along the parabolic profile (see Fig.~\ref{fig1}b)~\cite{Bao2012}:
\begin{equation}
p_\mathrm{el}(\xi)=\frac{V_\mathrm{g}^2}{2\varepsilon_0}\int\limits_{-L/2}^{+L/2} \! c^2\left(x\right) \,\,\frac{\mathrm{d}x}{L},
\label{eq:p}
\end{equation}
with $c(x)=(c_\mathrm{0}^{-1}(x)+c_\mathrm{SiO_2}^{-1})^{-1}$, $c_\mathrm{SiO_2}=\varepsilon_0\varepsilon_{\rm SiO{_2}}/(h_{\rm SiO_2})$ and $c_\mathrm{0}(x)=\varepsilon_0/(h(x))$, where $h(x)$ is the distance between the SiO$_2$ layer and the membrane, $\varepsilon_0$ is the vacuum permittivity and $\epsilon_{\rm SiO_2}\approx 3.9$ is the DC dielectric constant of SiO$_2$. We can thus compute $\xi$ as a function of $V_{\rm g}$ by injecting Eq.~\eqref{eq:p} into Eq.~\eqref{eq:pz} and solving the resulting equation numerically.

As shown in Fig.~\ref{fig3}a, our experimental measurements of $\xi$ vs $V_{\rm g}$ are very well fit to Eq.~\eqref{eq:pz}, using $E=1.05$\,TPa, $t=0.335$\,nm and $\nu=0.16$ \cite{Bacon1951,Lee2008,Jiang2009,Metten2014} and leaving $T_0=0.25\pm 0.03 \: \rm N/m$ as the only fitting parameter. This value corresponds to a built-in tension of $0.07 \pm 0.01 \%$, which is consistent with previous studies of suspended graphene monolayers~\cite{Bunch2007,Chen2009,Lee2008,Huang2011,Metten2013}. Let us note that the good agreement between our data and our model using an \textit{intrinsic} value of $E$~\cite{Lee2008,Bao2012,Koenig2011,Metten2014,Jiang2009} suggests that crumpling in our mechanically exfoliated graphene membrane can presumably be neglected~\cite{Nicholl2015}. 

Interestingly, Eq.~\eqref{eq:pz} has no real solution above a critical gate voltage $V_\mathrm{g,c}$. In particular, our model predicts an abrupt increase of $\xi$ for $V_\mathrm{g}>+20$\,V up to $+23.5$\,V, consistent with the collapse of the membrane observed for $V_{\rm g}>+19~\rm V$. We rely on the good accordance of our data with the electromechanical model to calculate the gate-induced deflection for a range of $h_0$  and compute the critical voltage $V_\mathrm{g,c}$ above which no real solution is found. The resulting contour plot is shown in Fig.~\ref{fig3}b for $L=5~\rm \mu m$ and an initial SiO$_2$ thickess  of $500~\rm nm$. Interestingly, for $h_0 < 180~\rm nm$, we find that increasing $V_{\rm g}$ leads to a smooth deposition of the graphene  layer onto the underlying SiO$_2$ substrate (\textit{i.e.,} the critical deflection $\xi_{\rm c}=h_0$), whereas for $h_0 > 180~\rm nm$ (as is the case in our experiment) the membrane abruptly collapses after reaching a value $\xi_{\rm c}<h_0$. Now, another interesting question for device optimization is, which voltage does a membrane sustain (oxide breakdown notwithstanding) until the membrane collapses? In Fig.~\ref{fig3}c, the critical lines ($\xi_{\rm c} ~vs ~ V_{\rm g,c}$) are shown  for different $L$, with color-coded $h_0$. In order to estimate the impact of a typical built-in tension, the critical lines for $L=3$ and $5$\,$\mu$m are added for $T_0=0.3$\,N/m. A finite built-in tension slightly shifts the curves to higher $V_\mathrm{g,c}$ and retards the deflection and the collapse.  Similar simulations for trenches etched in a Si/SiO$_2$ substrate with initial SiO$_2$ thicknesses of $285~\rm nm$ and $90~\rm nm$ are reported in the Supporting Information.

\begin{figure*}[!t]
\begin{center}
\includegraphics[width=1\linewidth]{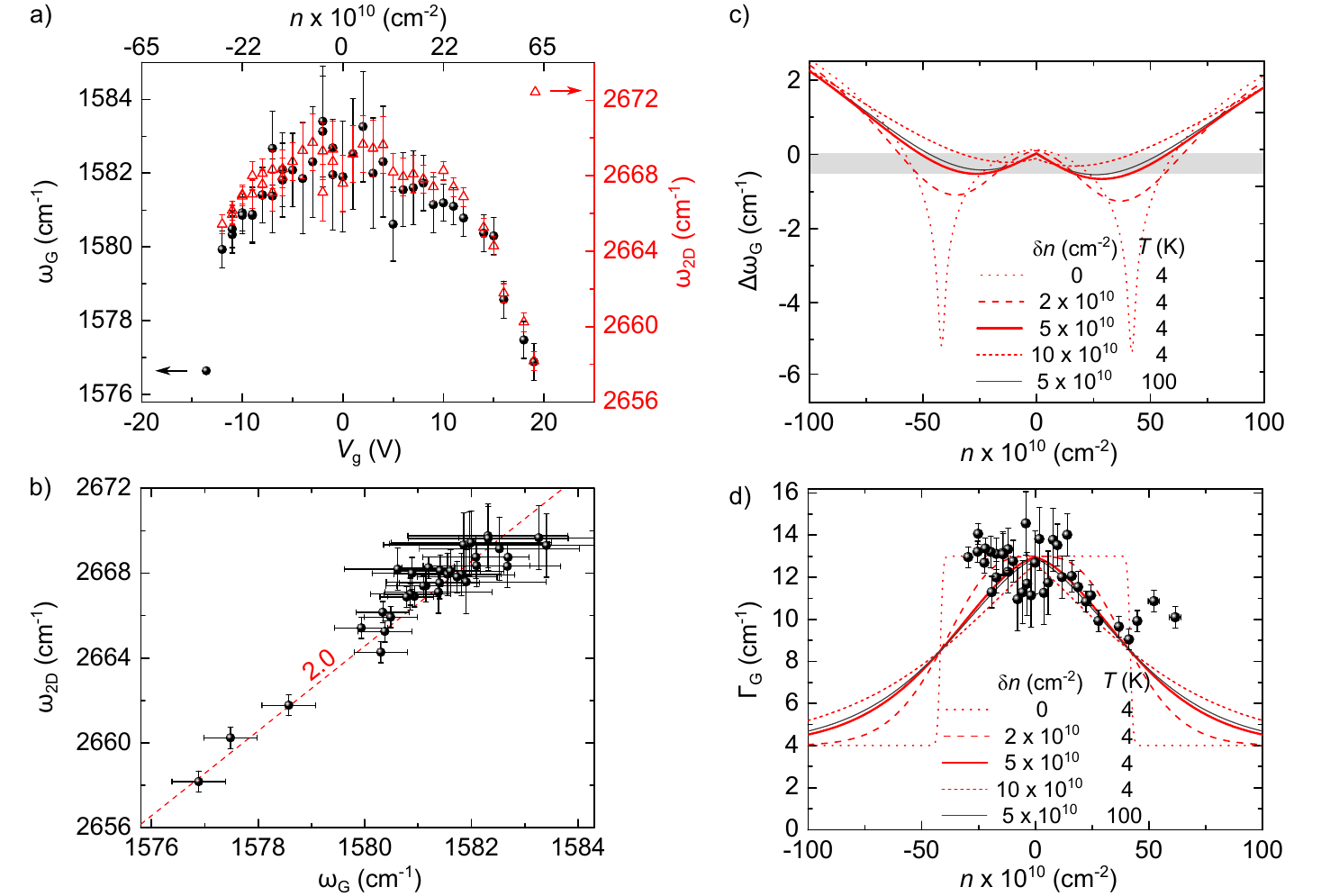}
\caption{(a) $\omega_\mathrm{G}$ (black spheres) and $\omega_\mathrm{2D}$ (red triangles) as a function of ~$V_\mathrm{g}$. (b) Correlation between $\omega_\mathrm{2D}$ and $\omega_\mathrm{G}$. The red line is a linear curve with a slope of 2.0. (c) Theoretically predicted shift of the Raman G-mode frequency ($\Delta \omega_{\rm G}$) and (d) full width at half maximum $\Gamma_{\rm G}$ due to electron-phonon coupling, calculated as in Refs.~\cite{Lazzeri2006,Pisana2007}. Calculations are performed as a function of the charge carrier density $n$ for $\rm T=4~K$ (red lines) and $\rm T=100~K$ (thin gray line), using a Fermi velocity of $1.3\times 10^6 ~\rm m/s$~\cite{Elias2011,Faugeras2015}, an electron-phonon coupling constant of $3.6\times 10^{-3}$, and considering various levels of residual charge inhomogeneity $\delta n$, as in Ref.~\cite{Froehlicher2015}. Our experimental measurements of $\Gamma_{\rm G}$ (black symbols) are compared to calculations in (d). The gray bar in (c) illustrates the very small G-mode frequency shifts due to electron-phonon coupling predicted for a realistic value of $\delta n=5\times 10^{10}~\rm cm^{-2}$. A residual contribution of $4~\rm cm^{-1}$ due to our spectral resolution and phonon anharmonicities~\cite{Bonini2007} is added to $\Gamma_{\rm G}$ in (d).}
\label{fig4}
\end{center}
\end{figure*}

\noindent\textbf{Electrostatically-induced strain and doping} 
The strength of our Raman scattering-based study lies not only in the \textit{in situ} determination of the membrane deflection, but also in the possibility of simultaneously extract local information about strain and doping, which is encoded in the Raman frequencies and linewidths. In Fig.~\ref{fig4}a, we show the evolution of $\omega_\mathrm{G}$ and $\omega_\mathrm{2D}$ with $V_\mathrm{g}$, extracted from the spectra in Fig.~\ref{fig1}c. The maximum Raman frequencies at $V_\mathrm{g}=0$ are $1582.6$\,cm$^{-1}$ and $2669.4$\,cm$^{-1}$, respectively. Both frequencies downshift with increasing $V_\mathrm{g}$ and a linear correlation is observed between $\omega_\mathrm{2D}$ and $\omega_\mathrm{G}$, with a slope of $2.0\pm0.2$ similar to previously observed values in strained graphene~\cite{Huang2009,Mohiuddin2009,Zabel2011,Lee2012a,Metten2013,Metten2014,Androulidakis2015} (see Fig.~\ref{fig4}b). However, the gate bias also dopes graphene.  A $T=4$\,K, as the Fermi energy of graphene approaches half the G-mode phonon energy~\cite{Lazzeri2006,Ando2006,Yan2008}, one might also expect significant anomalous G-mode softening, which could lead to a non-linear correlation between $\omega_\mathrm{2D}$ and $\omega_\mathrm{G}$ (Fig.~\ref{fig4}c).

In our study, the highest $n$ attained near device collapse is $\approx 6\times 10^{11}$\,cm$^{-2}$. Assuming a Fermi velocity $v_{\rm F}=1.3\times 10^6~\rm m/s$ in suspended graphene~\cite{Elias2011,Faugeras2015}, this value corresponds to a Fermi energy shift close to $\hbar\omega_{\rm G}/2$. As can be seen in Fig.~\ref{fig4}d, the gate-induced G-mode softening is also accompanied by a reduction of $\Gamma_{\rm G}$ by $\approx 3~\rm cm^{-1}$. This line narrowing cannot be explained by tensile strain and is, however, expected due to a reduction of Landau damping in doped graphene~\cite{Ando2006,Lazzeri2006,Yan2007,Pisana2007,Froehlicher2015}. Nevertheless, the smooth decrease of $\Gamma_{\rm G}$ with increasing $n$ observed here contrasts with the  ``flat hat'' behavior theoretically expected at $4~\rm K$, in the absence of charge inhomogeneity (see Fig.~\ref{fig4}d). This discrepancy can be rationalized by considering a small residual charge inhomogeneity on the order of $5\times 10^{10}~\rm cm^{-1}$~\footnote{Let us note that laser-induced heating is unlikely to play a role since we find very similar variations of $\omega_{\rm G}$ and $\Gamma_{\rm G}$ due to electron-phonon coupling at 4~K and 100~K (see Fig.~\ref{fig4}c), a temperature that is well above the unavoidable laser heating we can estimate in our conditions.}. In such conditions, the G-mode frequency is not expected to vary significantly with the gate-induced doping (see gray bar in Fig.~\ref{fig4}c). We also note that the 2D-mode frequency is expected to be virtually independent of the doping level for $\abs{n}<10^{12}~\rm cm^{-2}$~\cite{Yan2007,Froehlicher2015}.

We thus conclude that the G- and 2D-mode softening are essentially due to electrostatically-induced tensile strain, with a possible minor contribution from anomalous G-mode softening that may lead to a linear correlation with a slope slightly smaller that typically reported values~\cite{Huang2009,Mohiuddin2009,Zabel2011,Lee2012a,Metten2014,Androulidakis2015}. Our results demonstrate that the elusive phonon anomalies~\cite{Yan2008} will be intrinsically challenging to unveil in monolayer graphene as their fingerprints will be smeared out by tiny residual charge inhomogeneity and strain-induced phonon softening. However, the impact of the charge carrier concentration can still be probed through the narrowing of the G-mode feature (see Fig.~\ref{fig4}d). With the maximum G-mode frequency downshift of $\approx -5.5$\,cm$^{-1}$ with respect to its initial value at $V_\mathrm{g}=0$, we estimate an electrostatically-induced strain of $\sim 0.15\: (0.10)$\,\% in the case of uniaxial (biaxial) stress~\cite{Mohiuddin2009,Zabel2011,Kitt2013,Metten2014,Androulidakis2015,Polyzos2015}. This \textit{local} strain evaluated  through phonon softening in the middle of the trench is in qualitative agreement with an estimated \textit{average} strain of $0.3\pm 0.1\:\%$ based on the measurement of $\xi$ and assuming a parabolic profile. These estimations suggest that the electrostatically-applied stress is predominantly uniaxial, as can be expected within our sample geometry (see Fig.~\ref{fig1}a). The difference between the local and average strains indicates that strain near the edges of the trench is significantly larger than in the middle of the trench~\cite{Bao2012}. Near collapse, tensile strain may lead to a slight broadening of the G-mode feature ($\sim 1-2~\rm cm^{-1}$) that may partly compensate the suppression of Landau damping at large gate bias (see Fig.~\ref{fig4}d for $n \gtrsim 5\times 10^{11}~\rm cm^{-2} $). 

\noindent\textbf{Conclusion} 
Using Raman spectroscopy, we have achieved a contactless, minimally invasive  study of electrostatically-gated suspended graphene. The membrane deflection is estimated with a vertical resolution as low as a few nm and a diffraction-limited lateral resolution. A simple electromechanical model provides a faithful description of the membrane deflection, up to irreversible collapse, and an estimation of the built-in tension, a poorly controlled parameter, however of utmost importance in nanoresonators. Importantly, we precisely identify and separate the contributions of strain and doping in the Raman spectrum of suspended graphene. Accurate monitoring and  control of the deflection, stain and doping in suspended graphene and related atomically-thin materials is appealing not only for device engineering but also for opto-(electro)mechanical investigations~\cite{Reserbat2012,Barton2012,Reserbat2016,Davidovikj2016,Alba2016,Mathew2016,Schwarz2016}. 

\vspace{0.3 cm}



\noindent\textbf{Acknowledgement} 
We thank F.\,Federspiel, K.\,Makles, and P.\,Verlot for fruitful discussions, F.\,Chevrier, A.\,Boulard, M.\,Romeo, and F. Godel for experimental support, R.\,Bernard, S.\,Siegwald, and H.\,Majjad for assistance in the STNano clean room facility. We acknowledge financial support from C'Nano GE, the Agence Nationale de Recherche (ANR) under grants QuandDoGra 12 JS10-00101 and H2DH ANR-15-CE24-0016 and the University of Strasbourg Institute for Advanced Study (USIAS, GOLEM project). 




%


\onecolumngrid
\newpage
\begin{center}
{\LARGE\textbf{Supporting Information}}
\end{center}

\setcounter{equation}{0}
\setcounter{figure}{0}
\setcounter{section}{0}
\renewcommand{\theequation}{S\arabic{equation}}
\renewcommand{\thefigure}{S\arabic{figure}}
\renewcommand{\thesection}{SI\,\arabic{section}}
\linespread{1.4}

\bigskip

\section{Fitting procedures}
\label{S1}
The G-mode feature is fit to a Lorentzian profile and its spectral position $\omega_\mathrm{G}$, FWHM $\Gamma_\mathrm{G}$ and integrated intensity $I_\mathrm{G}$ are extracted. The 2D-mode feature shows a slight asymmetry, as often observed on clean suspended graphene devices~\cite{Berciaud2009,Lee2012,Luo2012,Berciaud2013}. Here, following Refs.~\cite{Basko2008,Venezuela2011,Berciaud2013}, we phenomenologically use the sum of two Lorentzian profiles at the power 3/2 with a shared FWHM in order to fit the 2D-mode feature and identify two subfeatures located at $\omega_\mathrm{2D_-}$ and $\omega_\mathrm{2D_+}$. The  low-energy feature is more intense (here, $I_\mathrm{2D_-}/I_\mathrm{2D_+}$ is $2.5\pm 0.3$) and $\omega_\mathrm{2D_+}-\omega_\mathrm{2D_-}$ is constantly $(12\pm 0.5)$\,cm$^{-1}$). In this Letter, we use $\omega_\mathrm{2D_-}$ to define the 2D-mode frequency $\omega_\mathrm{2D}$.
\section{Raman characterization at room temperature}
\label{S2}

\begin{figure}[htpb]
\begin{center}
\includegraphics[width=\linewidth]{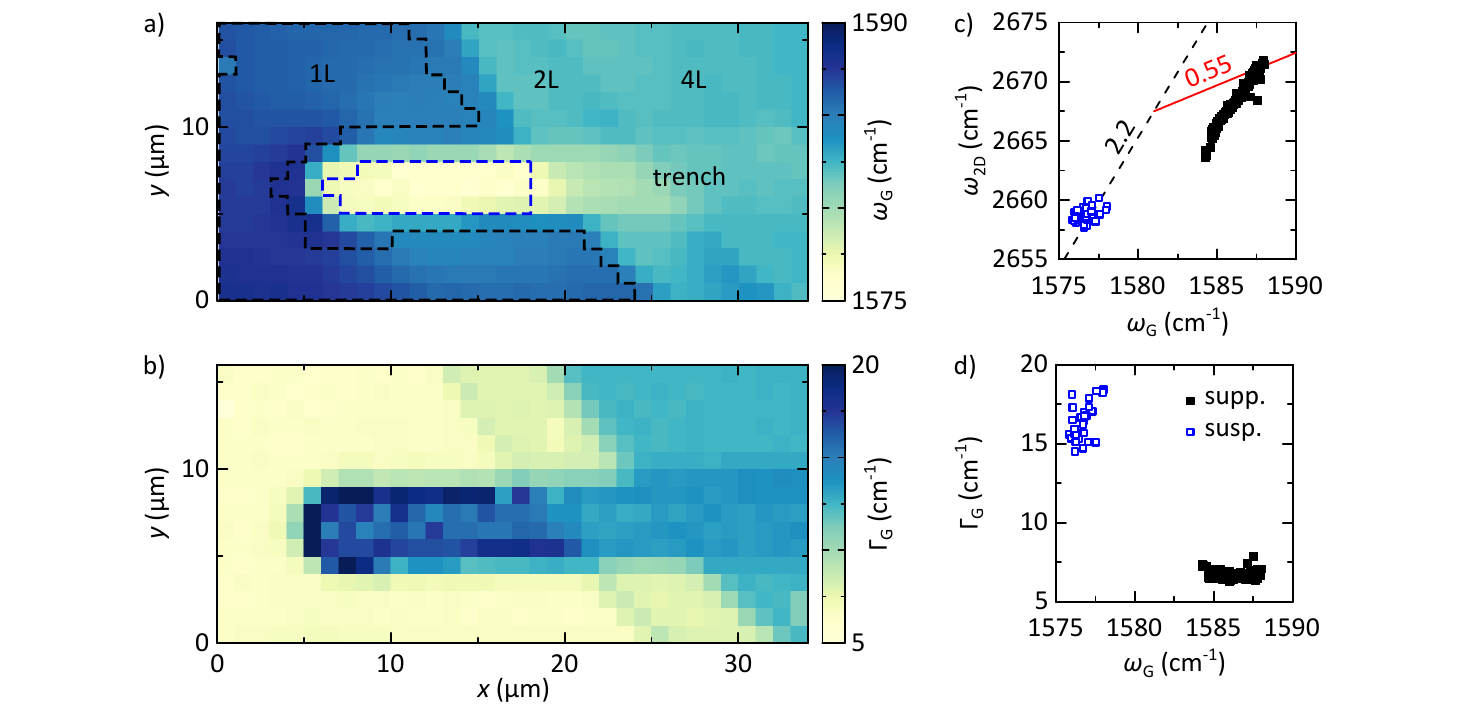}
\caption{(a) and (b) Room temperature hyperspectral Raman maps of the device, with a step size of $1$\,$\mu$m, showing the distribution of $\omega_\mathrm{G}$ and $\Gamma_\mathrm{G}$, respectively. The data taken into account for further analysis in (c) and (d) are delimited by a black (blue) dashed line in (a), for the supported (suspended) area. The suspended area is well distinguishable and characterized by a relatively low $\omega_\mathrm{G}$ and high $\Gamma_\mathrm{G}$. (c) Correlation of $\omega_\mathrm{G}$ and $\omega_\mathrm{2D}$ using the data extracted from the maps in (a) and (b). Data of the suspended (supported) area [see dashed lines in (a)] are shown as open blue (filled black) squares. A slope of $\partial \omega_\mathrm{2D}/\partial \omega_\mathrm{G}=2.2$ for strain \cite{Zabel2011,Lee2012a,Metten2014} and $0.55$ for hole doping \cite{Das2008,Froehlicher2015} are added as dashed black and straight red lines, respectively.}
\label{fig:SI1}
\end{center}
\end{figure}

Before our low-temperature study, the suspended graphene device has been characterized at room temperature by means of Raman spectroscopy using a laser beam at 532~nm. The G- and 2D-mode features are fit with the above-described procedure. The frequencies $\omega_\mathrm{G}$, $\omega_\mathrm{2D}$ and the linewidth $\Gamma_\mathrm{G}$ are extracted. Figures \ref{fig:SI1}a and b show Raman maps of $\omega_\mathrm{G}$ and $\Gamma_\mathrm{G}$.

 The graphs in Fig. \ref{fig:SI1}c and d correlate $\omega_\mathrm{G}$ with $\omega_\mathrm{2D}$ and $\Gamma_\mathrm{G}$, respectively. The data are extracted from the delimited areas on the map in Fig. \ref{fig:SI1}a and plotted as open blue squares for the suspended area and filled black squares for the supported area. On Fig. \ref{fig:SI1}c the data are concentrated with a small standard deviation and line up along a slope of $2.2$. Here we use the reference point ($\omega_\mathrm{G}^0=1581$\,cm$^{-1}$, $\omega_\mathrm{2D}^0=2667$\,cm$^{-1}$), as in ref.~\cite{Metten2013} for neutral, unstrained suspended graphene at $300$\,K and for a laser wavelength of 532~nm. In contrast, the data of the supported area are aligned along a line with the same slope of 2.2 but shifted to a higher $\omega_\mathrm{G}$ due to substrate-induced doping. Following the vector decomposition model proposed by Lee \textit{et al.}~\cite{Lee2012a} and applied to our suspended sample, we might estimate an average tensile strain of $0.06\pm 0.01$\,\% and a doping level  $\lesssim 10^{11}$\,cm$^{-2}$. The latter is in stark contrast with the doping level of the supported part, which is of one order of magnitude larger. The data represented in Fig.~\ref{fig:SI1}d combine the two maps in Fig.~\ref{fig:SI1}a and b, and confirm the quasi-undoped character of the membrane by the two well-distinguishable areas in the $\omega_\mathrm{G}$-$\Gamma_\mathrm{G}$-plane \cite{Lazzeri2006,Berciaud2009}.


\section{Raman enhancement due to optical interferences}
\label{S3}
The trench on which graphene is exfoliated and suspended can be considered as a multilayered system, \textit{i.e.} [Si - SiO$_2$ - air/vacuum - graphene - air/vacuum]. The involved layer thicknesses are on the same order of magnitude as the wavelength of the light, so that optical interference effects come into play. We model the evolution of the ratio of the incoming and scattered electrical field with a Fabry-P\'erot calculation, according to

\begin{equation}
A_{j+1,j}=\frac{r_{j+1,j}+A_{j,j-1}e^{2i\phi_j}}{1+r_{j+1,j}A_{j,j-1}e^{2i\phi_j}}.
\label{eq:Aj+1,j}
\end{equation}

Here, $r_{ij}$ are the reflection coefficients at the $i/j$ interface and the phase is defined as $\phi_i=2\pi \tilde{n}_id_i/\lambda$. The Raman interaction takes place in the last (graphene) layer and has to be considered separately. Multiple reflections within the graphene layer for the incoming (laser) and Raman-scattered light are calculated by taking the effective reflection coefficient (on one side of the graphene membrane). An optical enhancement factor is then obtained by integrating over the thickness $t$ of the membrane \cite{Blake2007,Yoon2009} according to

\begin{equation}
\mathrm{Enhancement\,factor}=\int\limits_{0}^{t}\!\left|\frac{E_x}{E_0}\times\frac{E_\mathrm{out}}{E_R}\right|^2\,\mathrm{d}x,
\label{eq:F}
\end{equation}

where $E_0$ is the electric field of the incoming laser beam, $E_x$ the electric field of the incoming laser beam at the position $x$ where the Raman interaction takes place, $E_\mathrm{R}$ the Raman electric field created at $x$ and $E_\mathrm{out}$ the Raman electric field coming out of the layered system. The enhancement factor is generally normalized relative to the case of monolayer graphene in free space. Here, for a clearer comparison with our experimental data (see Fig. 2a and 2b in the main text), we renormalize the amplitude of the enhancement factor variations (between its minimal and maximal values) by the measured ratio between the minimal and maximal integrated Raman intensities we have observed experimentally. This renormalization has no significant consequence on the determination of the membrane deflection.


\section{Details on the electromechanical model}
\label{S4}
The deflection  is modeled by regarding the electrostatic pressure on the membrane, which writes

\begin{equation}
	p_\mathrm{el}\left(V_\mathrm{g},\xi(x)\right)=\frac{\varepsilon_0\varepsilon_{\mathrm{SiO_2}}^2V_\mathrm{g}^2}{2\left(h_\mathrm{SiO_2}+\varepsilon_{\mathrm{SiO_2}}(h_0-\xi(x))\right)^2}.
	\label{eq:p1SI}
\end{equation}

$\xi(x)$ is the deflection profile perpendicular to the trench ($x$ direction), $\varepsilon_0$ the dielectric constant, $\varepsilon_\mathrm{SiO_2}$ the relative dielectric constant of SiO$_2$ and $h_\mathrm{SiO_2}$ the residual oxide thickness.

The membrane deflection in the middle of the trench (simply denoted $\xi$ in the main manuscript), induced by a uniform pressure load writes \cite{Medvedyeva2011,Bao2012}

\begin{equation}
p_\mathrm{el}(\xi)=\frac{64Et}{3(1-\nu^2)L^4}\,\xi^3+\frac{8T_0}{L^2}\,\xi,
	\label{eq:pzSI}
\end{equation}

where $E$ is the Young's modulus, $\nu$ the Poisson ratio and $t$ the thickness of the membrane, $L$ the width of the trench (see Fig. 1b in the main text) and $T_0$ its pre-strain (in N/m). As mentioned in the main text, the scaling $p_\mathrm{el}\propto\xi^3$ is well-known for membranes (without bending rigidity) \cite{Hencky1915,Komaragiri2005,Yue2012,Metten2014}, and the additional pre-strain-term can be interpreted as an effective bending rigidity moderating the deflection, as $p_\mathrm{el}$ scales linearly with $\xi$ for thin plates \cite{Landau1970,Vlassak1992,Beams1995,Williams1997,Medvedyeva2011}.

Eq.~\eqref{eq:pzSI} is obtained from the equation of equilibrium \cite{Landau1970},

\begin{equation}
p_\mathrm{el}(\xi(x))=D\,\frac{\partial^4\xi(x)}{\partial x^4}-T(\xi'(x))\,\frac{\partial^2\xi(x)}{\partial x^2},
\label{eq:equil}
\end{equation}

by supposing a parabolic profile of the membrane, and where $T=T_\mathrm{e}+T_0$ is the total strain in the membrane, with $T_\mathrm{e}$ the strain due to the electrostatically induced deflection. The profile is taken as $\xi(x)=p_\mathrm{el}(L^2/4-x^2)/(2T)$, which is a first approximation, because back coupling mechanisms between the deflection and the charge redistribution within the membrane are neglected. However, Eq.~\eqref{eq:equil} can be self-consistently solved considering a parabolic profile.

As discussed in the main text, Eq.~\eqref{eq:pzSI} requires a uniform pressure load over the membrane, which is calculated as an average value of $p_\mathrm{el}$, obtained by integrating the local capacitance $c$ along the parabolic profile $\xi(x)$:
\begin{equation}
p_\mathrm{el}(\xi)=\frac{V_\mathrm{g}^2}{2\varepsilon_0}\int\limits_{-L/2}^{+L/2} \! c^2\left(x\right) \,\,\frac{\mathrm{d}x}{L},
\label{eq:pSI}
\end{equation}
with $c(x)=(c_\mathrm{0}^{-1}(x)+c_\mathrm{SiO_2}^{-1})^{-1}$, $c_\mathrm{SiO_2}=\varepsilon_0\varepsilon_{\rm SiO{_2}}/(h_{\rm SiO_2})$ and $c_\mathrm{0}(x)=\varepsilon_0/(h_0-\xi(x))$  the distance between the SiO$_2$ layer and the membrane, $\varepsilon_0$ is the vacuum permittivity and $\epsilon_{\rm SiO_2}\approx 3.9$ is the DC dielectric constant of SiO$_2$. We can thus compute $\xi$ as a function of $V_{\rm g}$ by injecting Eq.~\eqref{eq:pSI} into Eq.~\eqref{eq:pzSI}:

\begin{equation}
V_\mathrm{g}^2=\frac{\frac{4\varepsilon_0}{c_\mathrm{SiO_2}^2}\left(\frac{8T_0}{L^2}\,\,\xi+\frac{64Et}{3(1-\nu^2)L^4}\,\,\xi^3\right)}{\sqrt{\frac{\varepsilon_0}{B^3c_\mathrm{SiO_2}\xi}}\,\,\mathrm{arctan}\left(\sqrt{\frac{c_\mathrm{SiO_2}\xi}{\varepsilon_0B}}\right)+\frac{1}{B^2+\frac{c_\mathrm{SiO_2}\xi}{\varepsilon_0}B}}.
\label{eq:case3}
\end{equation}
Here, $B=1+c_\mathrm{SiO_2}\,\left(h_0-\xi\right)/\varepsilon_0$ and the pre-strain $T_0$ can be used as a fitting parameter, as shown in Fig. 3a in the main text. Equation~\ref{eq:case3} is solved numerically.

\clearpage

\section{Deflection extracted from $I_\mathrm{2D}$}
\label{S5}

\begin{figure}[htpb]
\begin{center}
\includegraphics[scale=1.0]{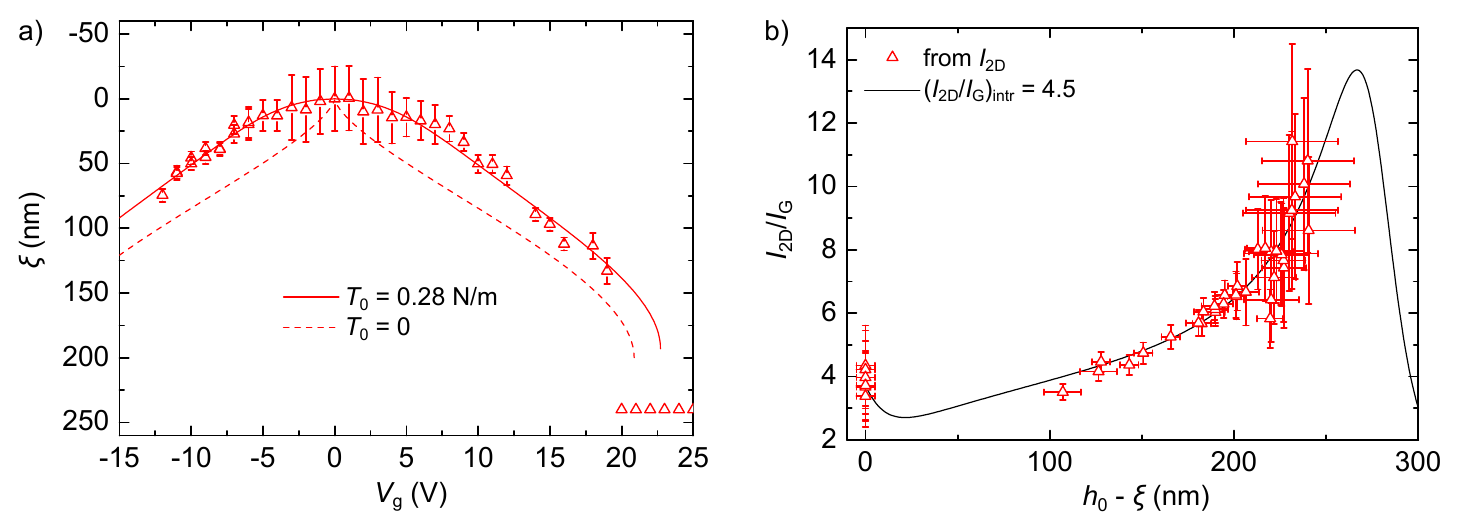}
\caption{Vertical deflection $\xi$ in the middle of the trench, determined by the variation of $I_\mathrm{2D}$, as a function of $V_\mathrm{g}$. The dashed red line is the electromechnical model with $T_0=0$ whereas the solid line is with $T_0$ as fitting parameter ($T_0=0.28$\,N/m). (b) Evolution of $I_\mathrm{2D}/I_\mathrm{G}$ as a function of $h_0-\xi$, determined from the evolution of $I_\mathrm{2D}$ with $V_\mathrm{g}$.  The red line is a fit based on a multiple interference model using  an \textit{interference-free} reference value of $(I_\mathrm{2D}/I_\mathrm{G})_\mathrm{intr}=4.5$, much similar to the value proposed in the main text (see also Fig. 2c).}
\label{fig:SI3}
\end{center}
\end{figure}

The values of $\xi$  presented in Fig 3a of the main manuscript are derived from the integrated intensity of the G-mode feature $I_{\rm G}$. In Fig.~\ref{fig:SI3} we show the data extracted from $I_\mathrm{2D}$ and the corresponding fit using our electromechanical model. A pre-strain of $T_0=0.28$\,N/m is found, which is comparable the one extracted from $I_\mathrm{G}$.

\clearpage

\section{Electromechanical calculations for various SiO$_2$ thickness}
\label{S6}

\begin{figure}[htpb]
\begin{center}
\includegraphics[scale=1.0]{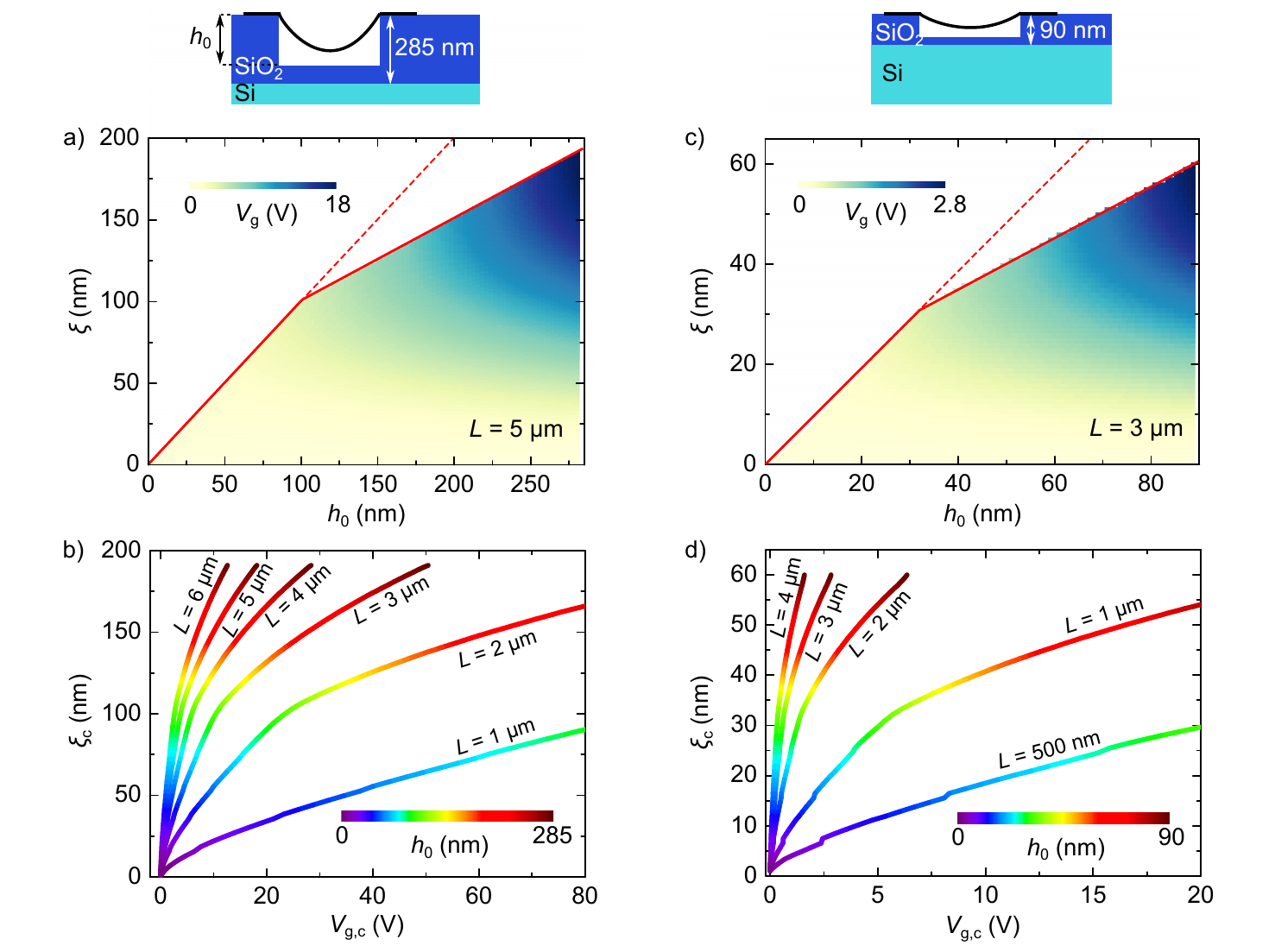}
\caption{(a) False color map of the gate voltage required to attain a given $\xi$ as a function of $h_0$, for $L=5$\,$\mu$m and an initial SiO$_2$ thickness of 285 nm [see schematic above (a)]. (b) Critical deflection $\xi_\mathrm{c}$ as a function of the corresponding critical gate voltage $V_\mathrm{g,c}$. The data are extracted from the color plots, as explained in the main text. The critical lines are shown for different trench widths $L$. The color code indicates the corresponding $h_0$. (c) False color map as in (a), plotted for an initial SiO$_2$ thickness of $90$\,nm and a trench width of $L=3$\,$\mu$m. (d) As in (b), the critical deflection $\xi_\mathrm{c}$ is plotted as a function of $V_\mathrm{g,c}$ and $h_0$. The chosen values of $L$ are adapted to the smaller values of $h_0$.}
\label{fig:SI5}
\end{center}
\end{figure}

Commonly used Si/SiO$_2$ substrates have oxide thickness of $285$ and $90$\,nm. In Fig.~\ref{fig:SI5}, we show data similar to Fig. 3b-c in the main manuscript for for $L=5$\,$\mu$m and $L=3$\,$\mu$m, respectively.

\clearpage


\section{Critical charge carrier density}
\label{S7}

\begin{figure}[htpb]
\begin{center}
\includegraphics[scale=1.0]{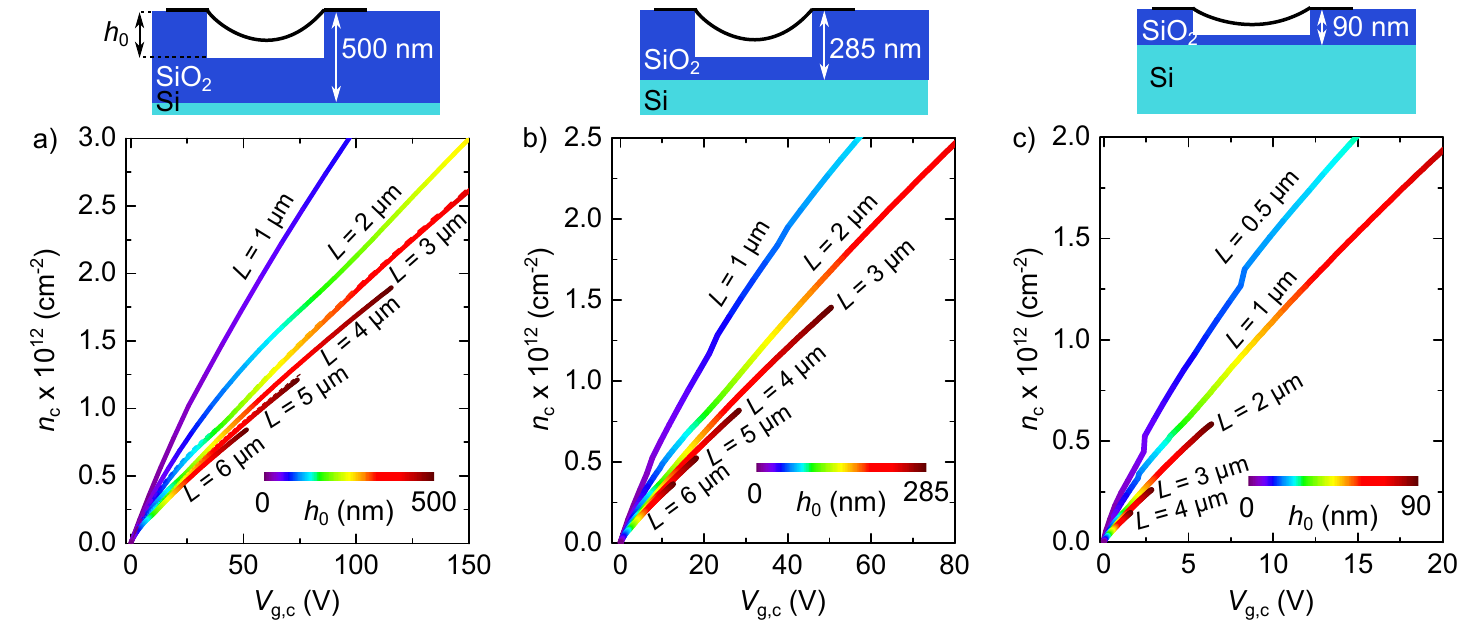}
\caption{The charge carrier density in the middle of the trench is calculated according to $n(V_\mathrm{g},\xi)=\frac{\varepsilon_0\varepsilon_{\mathrm{SiO_2}}V_g}{\varepsilon_{\mathrm{SiO_2}}(h_0-\xi)+h_\mathrm{SiO_2}}$. In (a), (b) and (c) the predicted critical charge carrier density $n_{\rm c}$ as a function of the critical gate voltage (near membrane collapse) and $h_0$ are shown for an initial oxide thickness of $500$, $285$ and $90$\,nm.}
\label{fig:SI7}
\end{center}
\end{figure}

\clearpage

\section{Scanning Electron Microscopy (SEM) imaging}
\label{S8}

\begin{figure}[htpb]
\begin{center}
\includegraphics[scale=0.92]{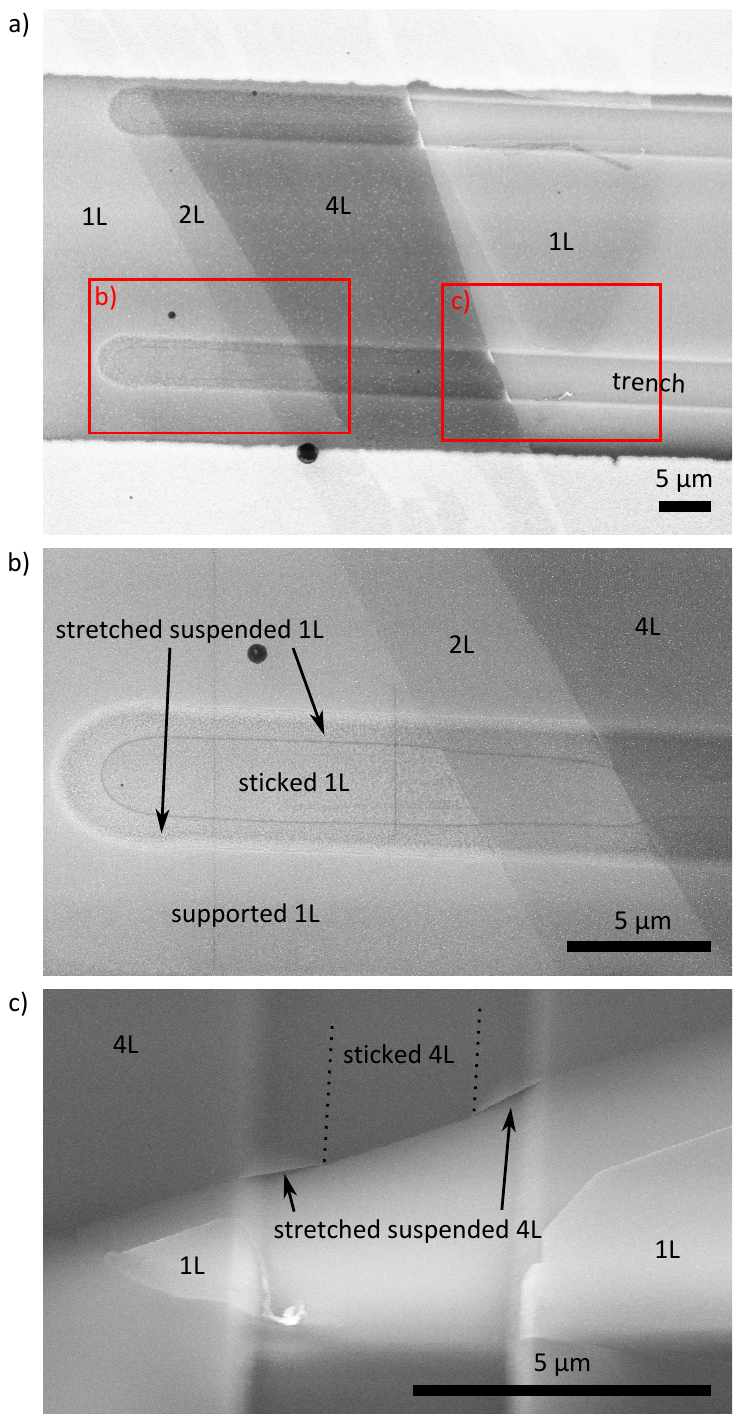}
\caption{SEM images of the device after the gate-dependent Raman study, after the membrane has collapsed and adheres to the bottom of the trench. (a) Overview of the device with red squares indicating the zoomed area in (b) and (c). (b) The collapsed area can be distinguished by a light gray line in the bottom. Also the neighboring 2LG and 4LG are stuck to the substrate. The stretched suspended part at the borders of the trench is seen in (c) for the 4LG, where the image is rotated by $90$ degrees and slightly tilted. The scale bar is $5$\,$\mu$m in all images.}
\label{fig:SI2}
\end{center}
\end{figure}

\end{document}